\documentclass{article}
\usepackage{color}
\usepackage{graphicx}

\begin{document}
\title{Deformation Lamps: A Projection Technique to Make a Static Object Dynamic}
\author{Takahiro Kawabe\thanks{kawabe.takahiro@lab.ntt.co.jp} \and Taiki Fukiage \and Masataka Sawayama \and Shin'ya Nishida}
\date{NTT Communication Science Laboratories, Nippon Telegraph and Telephone Corporation, Japan}

\maketitle
\begin{abstract}
Light projection is a powerful technique to edit appearances of objects in the real world. Based on pixel-wise modification of light transport, previous techniques have successfully modified static surface properties such as surface color, dynamic range, gloss and shading. Here, we propose an alternative light projection technique that adds a variety of illusory, yet realistic distortions to a wide range of static 2D and 3D projection targets. The key idea of our technique, named `Deformation Lamps', is to project only dynamic luminance information, which effectively activates the motion (and shape) processing in the visual system, while preserving the color and texture of the original object. Although the projected dynamic luminance information is spatially inconsistent with the color and texture of the target object, the observer's brain automatically combines these sensory signals in such a way as to correct the inconsistency across visual attributes. We conducted a psychophysical experiment to investigate the characteristics of the inconsistency correction, and found that the correction was dependent critically on the retinal magnitude of inconsistency. Another experiment showed that perceived magnitude of image deformation by our techniques was underestimated. The results ruled out the possibility that the effect by our technique stemmed simply from the physical change of object appearance by light projection. Finally, we discuss how our techniques can make the observers perceive a vivid and natural movement, deformation, or oscillation of a variety of static objects, including drawn pictures, printed photographs, sculptures with 3D shading, objects with natural textures including human bodies. 
\end{abstract}

\section{Introduction}

In spatial augmented reality, the user's physical environment is augmented with light-projected images that are integrated directly in the user's environment \cite{RWF1998}. The light projection can dramatically modify the appearance of a real object surrounding us by changing the light transport while keeping the object's physical material and shape intact. Earlier studies propose a technique which can add a virtual appearance to an object with matte uniform surfaces \cite{MNS2004,RWL2001}. Later studies developed image compensation methods which optically eliminate the effects of undesirable textures and pigments in non-optimal surfaces \cite{BEK2005,BIWG2008}. The compensation methods were extended to successful modification of the appearance of complexly textured surfaces with respect to color \cite{AOSS2006}, dynamic range \cite{BI2008}, gloss and shading \cite{AMANO2013}. 

Not only static aspects, but also dynamic aspects of the object appearance are subject to modification by light projection. For example, a previous study \cite{RZW2002} demonstrated a technique that indirectly adds motion impressions to a static object by projecting a moving pattern to the background and generating a visual illusion called `induced motion' \cite{D1929}. A recent study has developed projector-based illumination with precise image compensation for production of high-frequency details of dynamic facial expressions of physical avatars \cite{BBGIBG2013}. 

Although these projection techniques give motion impressions to a static object by replacing the original appearance with a new one, is it possible for light projection to add a motion impression to a static object without changing the original appearance of the object? In other words, can we move a real static object just by projecting light? A straightforward solution one might imagine is to reproduce the shifted version of the original colors/textures on the object's surface while compensating for the effects of the original colors/textures. However, perfect reproduction of the original appearance by light projection is quite challenging, if not impossible. Even onto a uniform screen, it will be hard to perfectly simulate the appearance of a real object by light projection due to the limitations in spatial resolution, dynamic range and color reproduction of the current projection system. In addition, one will be able to perfectly compensate for the effects of non-uniform textures of the object's surface only under restricted conditions: i.e., the lowest surface light reflectance should not be low in all color channels; the projection light should be intense; and the ambient illumination should be weak. These constraints will likely reduce the range of application, or change the object's appearance significantly different from the natural one. Therefore, a novel approach was warranted to animate static real objects in our daily environment while maintaining the original appearance of the objects.

Here, we propose a novel light projection technique named `Deformation Lamps' that makes a static object appear to deform and move with keeping the object's appearance almost intact. Put simply, Deformation Lamps only superimposes dynamic luminance signals onto a colorful static object and produces an illusory, but realistic movement of the object.  

\begin{figure}[]
  \centering
  \includegraphics[width=4.5in]{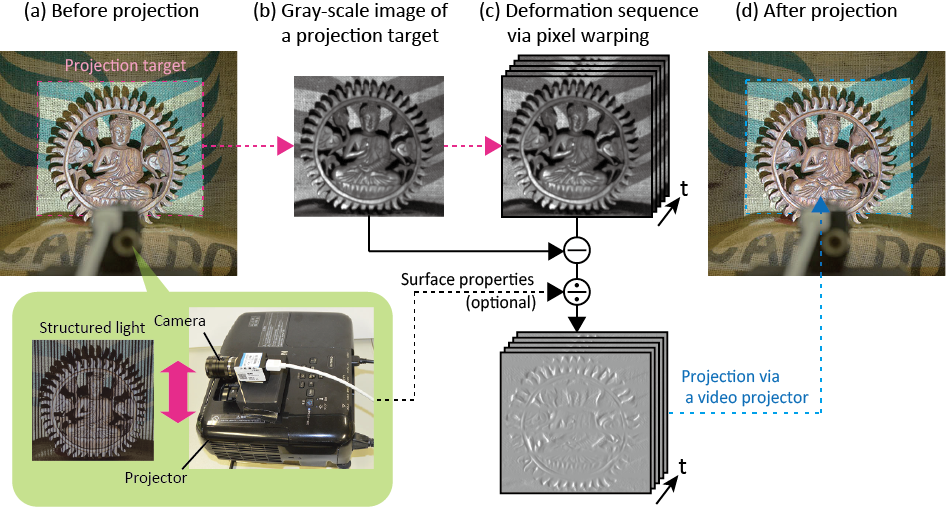}
  \caption{The processing stream of a representative system of Deformation Lamps, which consists of a video projector and a camera. (a) A static object is placed as a projection target. (b) A gray-scale camera image of the static object. (c) A deformation image sequence is generated by deforming the camera image (b) with an arbitrarily defined deformation map. (d) Difference image sequence obtained by subtracting (b) from (c). (e) The difference image sequence is optically projected onto the static object. Compensation of the projection image is optional. Humans perceive an illusory dynamic deformation of the static object. }
\end{figure}

A representative system of Deformation Lamps uses a camera-projector system (Figure 1). The system first takes a gray-scale image of a colorful target object through a camera  (Upper part in Figure 1b). Second, it creates a sequence of deformed gray-scale images by dynamically deforming the gray-scale image in accordance with a sequence of pre-defined deformation maps (Upper part in Figure 1c). Third, the system subtracts the pixel intensity value of the original image from each deformation image . This results in a sequence of intensity difference images produced by the deformation sequence (Lower part in Figure 1c). Finally, the sequence of intensity difference images is projected as a gray-scale movie onto the target object (Lower part in Figure 1d). Observers viewing the target object have an illusory percept as if it, including its color components, is moving or deforming. Compensation of the projection image is optional. By manipulating the deformation pattern, Deformation Lamps can make the objects in the picture appear to, say, sink in running water or flutter in the wind. With proper image alignments, the Deformation Lamps can also animate 3D textured/shaded objects.

According to the standard definition of image movement (i.e., spatial shifts of intensity/color pattern over time), Deformation Lamps does not always produce image movements of the intensity component in a physically correct way, nor does it produce image movements of the color component at all. Nevertheless, this technique is able to generate a vivid and natural appearance of colorful motion, thanks to the processing characteristics of the human visual system. Our visual system separately analyzes basic visual attributes such as motion, form, and color, and integrates them into coherent representations in a subsequent processing stage (e.g., \cite{LH1987}). Deformation Lamps adds dynamic components of image deformation (motion information together with dynamic form information) to a projection target that keeps the original undistorted static form and color information. As a result, the dynamic information and static information are physically inconsistent with each other. However, the human visual system attempts to resolve such consistency when it integrates information across attributes into coherent scene representations. For instance, this attribute integration is known to cause a visual illusion called `motion capture' wherein the position and shape of a static color signal is perceptually pulled in the direction of accompanying motion signals \cite{RAMA1987,RAMACAVA1987}. By exploiting such perceptual inconsistencies, Deformation Lamps is able to give the impression of natural movement to a static object. It should be also noted that image compensation of luminance component is helpful for accurate control of induced motion, but perfect compensation is not always best, since strong projection may alter the original object’s appearance. Image compensation is not critical at least for the purpose of giving a smooth apparent movement signals to human observers. Therefore, Deformation Lamps works well with a wide range of projection condition under natural ambient illumination.  

In what follows, we will first describe the relationship of Deformation Lamps with the previous light projection techniques and other related work. Then, we will describe in detail how Deformation Lamps works, and report psychophysical experiments that evaluated its performance. After describing possible applications, we will end by discussing its limitations and future issues.

\section{Related Work}

Deformation Lamps utilizes a video projector to give dynamic impressions to static objects. This technique is related to spatial augmented reality, wherein virtual objects are created in the real world without the viewer having to wear special devices \cite{BR2005,RWF1998}. Past research in this field has come up with a variety of projection methods \cite{AMANO2013,RWL2001,U1997}. Raskar and his colleagues proposed `Shader Lamps' \cite{RWL2001} that can change the appearances of real objects, including their color, texture, and material properties, into those of virtual objects. Bimber and Iwai \cite{BI2008} proposed a light projection method to enhance the luminance contrast and color of printed materials, and Bimber et al. and later studies have developed algorithms to modify the appearance of real objects by light projection with image compensation\cite{AMANO2013,BIWG2008}. The technique is used to edit the appearance of real objects \cite{ALY2008,AYL2012}, and moreover, is also employed to add motion impressions to a static object by projecting a moving pattern \cite{BBGIBG2013,RZW2002}.

Deformation Lamps produces apparent image movements not by shifting the position of the object image as conducted in the previous studies, but by adding luminance motion signals that activate the motion sensors in the human visual system. In support of our strategy, past studies have invented several dynamic displays that produce vivid motion sensations without the corresponding position shifts in the image. In the phenomenon known as `reversed phi' \cite{ANSTIS1970,AR1975}, the perceived motion direction of an object moving across two video frames is reversed when the luminance contrast polarity of the image is reversed. In a display entitled `Motion without movement' \cite{FAH1991}, local phase shifts of the luminance pattern produce the perception of a global motion flow in a direction consistent with the phase shifts. Even static pictures can produce illusory motion sensations when they activate the motion sensors of the human visual system (See, e.g., \cite{FW1979,KA2003}, and a graphic technique for automatically generating such illusory motion patterns has been proposed \cite{CLQW2008}.

Deformation Lamps can add a variety of distortions/movements to static objects. Without using light projection, several computer-graphics-based image editing techniques have been proposed in order to give motion impressions to a static image. \cite{C2005} reported that human-annotated objects such as water, boats, and clouds, could be animated by manipulating the static image on the basis of stochastic motion textures as a two-dimensional deformation map. More recently, \cite{BBBVN2011} edited liquid flow impressions in a static image by means of physically based animation that uses a ray tracing method supporting refraction and reflection.

\section{Basic algorithm}
Hereafter, we describe in detail how Deformation Lamps adds dynamic impressions to a static real object. 

The goal of this technique is to produce a dynamic image sequence that gives observers an impression of an object movement, from a combination of a static object and a projected movie. In general, a movie, $I_{movie}$, can be described as a linear combination of a static object and the residual dynamic component in the following way.

\begin{equation}
 I_{movie}(x,y,t) = I_{static}(x,y)+I_{dynamic}(x,y,t).
\end{equation}

If $I_{static}$ is the zero-temporal-frequency (i.e., temporally averaged) image of $I_{movie}$ (or an image close to it), then $I_{dynamic}$ is (approximately) the non-zero-temporal-frequency (i.e., pure dynamic) components of $I_{movie}$. When $I_{movie}$ contains color information, both $I_{static}$ and $I_{dynamic}$ usually contain color information. 

A key idea of Deformation Lamps is to produce an approximate color movie perceptually indistinguishable from a real one from a linear combination of a static color picture and a dynamic gray-scale movie. The approximated color movie on a display is described as follows,  

\begin{equation}	
I_{movie} (x,y,t) \approx I_{static}(x,y) +I_{luminance \; dynamic}(x,y,t),
\end{equation}
where
\begin{equation}	
I_{luminance \; dynamic} (x,y,t) = I_{luminance \; movie}(x,y,t) - I_{luminance \; static}(x,y).
\end{equation}

Here, we want to make the static color object perceptually dynamic by projecting luminance dynamic signals $P$ onto the object. For the sake of simplicity, we assume that the projection target is a color picture, which has a Lambertian surface with reflectance $K$. Under this assumption, a color movie approximated by using the light projection of motion signals onto the picture is described as follows,

\begin{equation}
 I_{movie} (x,y,t) \approx K(x,y)(Env(x,y,t) + P(x,y,t)) ,
\end{equation}

where $Env(x,y,t)$ denotes environmental ambient light. The projected light includes an arbitrary gray background $B$ so as not to take values below 0. Thus, $P$ is described as follows,

\begin{equation}
 P (x,y,t) =wI_{luminance \; dynamic}(x,y,t) + B ,
\end{equation}

where $w$ is a weight coefficient that modulates contrast of the dynamic component. In a case wherein users compensate luminance values in accordance with the reflectance of a picture, $I_{luminance \; dynamic}$ should be appropriately weighted (for example, w can be set as $\frac{1}{K}$). 

In the camera-projection system described in Figure 1, $I_{movie}$ is a deformation image sequence the system attempts to produce on the object. $I_{static}$ is the appearance of the projection target object under the projection of the gray background $B$. To get $I_{luminance \; dynamic}$, the camera first takes a gray-scale image of the object ($I_{luminance \; static}$) while the projector projects the gray background $B$. Second, the system spatiotemporally deforms it in accordance with a pre-defined deformation map sequence, {\it D}, to generate $I_{luminance \; movie}$. Specifically, this process uses a pixel warping method that solves the following equation,

\begin{equation}
	I(x',y',t_k)=T_{D(x,y,t_k)}[I(x,y)]
\end{equation}

where $T_{D(x,y,t_k)}$  is a spatial transform function using a deformation map at the kth frame, $D(x,y,t_k)$. In the actual implementation, we solved this equation as an inverse spatial transformation by using Matlab's {\it interp2} function. The system then compute $I_{luminance \; dynamic}$ by (3) and projects $P$ in (5) onto the static object to approxiate $I_{movie}$. The resulting object looks something akin to a movie $I_{movie}$ to human observers. 

Technically, the projected motion sequence and the static object need to be aligned. In general, manual alignment by the naked eye is sufficient for getting a reasonable visual effect as long as the projection targets are 2D pictures. For projection to 3D objects, precise image alignment using a structured light system \cite{LBJ2004} produced good visual effects. It was also effective to align the optical axis between the camera and projector \cite{AMANO2013}.

We found that simple projection of a dynamic achromatic pattern with manual adjustments of projection parameters such as contrast and gamma was sufficient to produce good illusory movements of real static objects in many cases. On the other hand, in order to exactly follow (2) to produce a quasi-chromatic movie, one could adjust the projected achromatic image using the known compensation methods \cite{AMANO2013}, which exclude the effects of spatial variation in reflectance (albedo) of the object’s surface, and match the contributions of the static object and the projected image to the final image formation. However, perfect compensation can be accomplished only under limited situations where the lowest surface reflectance is considerably high, and the ambient illumination is much weaker than the projected light. (Note that perfect compensation is even harder for chromatic images than for achromatic images, since the conditions should be met for all color channels). Even under the conditions where perfect compensation is theoretically possible, bright and high-contrast light projection will produce un-expected noise and a significant change in the appearance of the original static object. Therefore, perfect compensation is not only hard to obtain, but also non-ideal for Deformation Lamps. Fortunately, the technique produces good visual effects even when the projected luminance contrast is significantly lower than the point of perfect compensation, thanks to the characteristics of the human visual system (see below). It works even with partial compensation or without compensation. Deformation Lamps is robust against a wide variation in projection/illumination condition.

It should be also added that the spatial resolution of the projected image becomes lower than that of the static object due to optical blur.

\section{Perceptual process}
Given that Deformation Lamps produces an image sequence that only crudely approximates a movie, why can it produce a movie-like visual experience to human observers? The answer is related to the processing characteristics of human vision, particularly those of motion processing (see, e.g., \cite{BT2011,NISHIDA2011,WANDELL1995} for review).  

In the early stages, the human visual system analyzes image motion almost independently of its form and color processing \cite{CTF1984,LH1985,RAMAGRE1978}. The low-level motion mechanism is mainly driven by spatiotemporal image flow, or motion energy, of luminance modulation \cite{AB1985}. It does not require an exact pattern match over time \cite{ANSTIS1970,AR1975,FAH1991}. It has high contrast sensitivity to luminance motion energy at the low and middle ranges of spatial frequency \cite{K1980,WANDELL1995}. On the other hand, the low-level motion mechanism is insensitive to the static component of the luminance pattern \cite{LS1995,VS1985}. It is not very sensitive to chromatic motion signals \cite{CW2005,LH1985} or high-spatial frequency components \cite{K1980,WANDELL1995}, either. 

The characteristics of the human motion system explain why our technique can effectively drive human motion sensors even when the projected luminance pattern is at low contrast. A weighted summation of the original and shifted image results in a partially shifted image with the shift magnitude depending on the weight. Although this does not hold for high spatial frequency components (relative to the shift size), human motion sensors are sensitive the phase shift of lower bands of spatial frequency. Hence, when the contrast of a projected shifted pattern is reduced, the phase shift is reduced in the summation image, but it remains to be a natural shift. (One can compensate for the motion underestimation due to low contrast by exaggerating shifts in the projected image.) 

Given these properties of the low-level motion mechanism, we can consider that the gray-scale dynamic image component ($I_{luminance \; dynamic}$) contains almost all the motion information needed to drive this mechanism. Even when the dynamic component is mixed with the static component ($I_{static}$) through projection, the low-level motion mechanism will respond only to the dynamic component, and the relative weight of the dynamic component is not important as long as it is large enough to drive the mechanism that has high contrast sensitivity. In addition, the motion mechanism favoring low-frequency patterns is not affected much by the high-frequency reduction caused by optical blur. Therefore, in Deformation Lamps, the projected dynamic image component is expected to activate the low-level motion mechanism in almost the same manner as would a color movie that the system intended to produce (i.e., $I_{movie}$(x,y,t)).

What about color? Deformation Lamps affects the color distribution of the static target object only a little. Strictly speaking, specular reflection at the object surface could produce a mixture of object color and light color, but this color change is mainly in the saturation. What about form? Dynamic luminance form information (e.g., moving edges) is added to the object. At the same time, static luminance form information is left unerased. As such, Deformation Lamps does not produce correct movie signals for color and form processing. The final image includes inconsistencies between the dynamic information (motion and dynamic form) from the projector and the static information (color and static form) of the object.

\begin{figure*}[!t]
  \centering
  \includegraphics[width=4.5in]{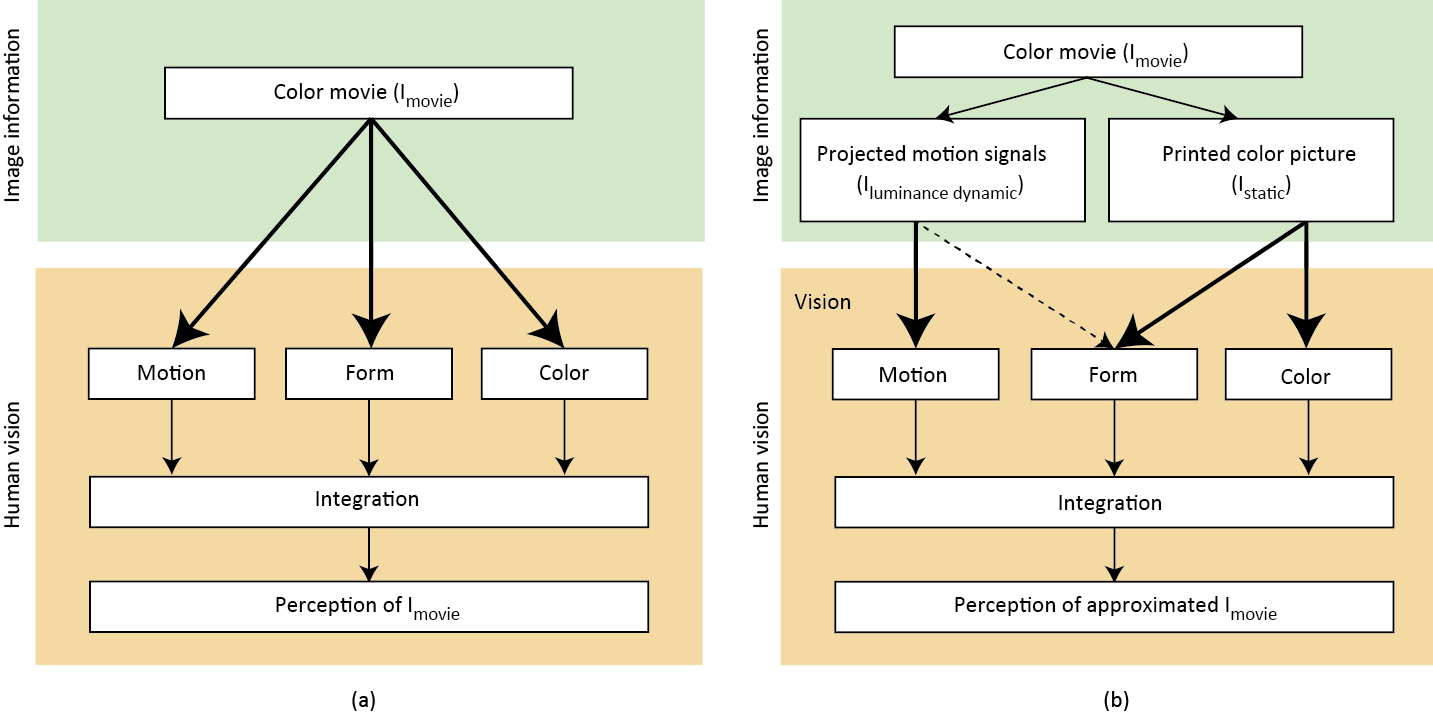}
  \caption{Image information and perceptual processing for (a) a color movie ($I_{movie}$) and (b) Deformaion Lamps ($I_{luminance \; dynamic}$ + $I_{static}$). Despite a significant difference in image formation, human vision produces similar perceptions as a result of separate analysis and sebsequent integration of motion, color and form.}
\end{figure*}

We consider that this inconsistency can be perceptually resolved by the brain when it integrates motion, form, and color signals (Figure 2). Because there is no moving object that lacks surface color and form in the real world, the human brain needs to integrate motion signals with color and form into coherent object representations. It is suggested that the brain assumes that the color and texture of a moving object should shift the position together with the motion signal and `corrects' the perception when this rule is violated. The evidence includes apparent movement of texture and color together with superimposed luminance motion (Motion Capture: \cite{RAMA1987,RAMACAVA1987}), as well as positional shifts of spatial patterns in the direction of motion \cite{RAMAAN1990,DD1991,NJ1999} and perceptual integration of form and color information across the spatiotemporal trajectory of motion \cite{NISHIDA2004,NWKT2007}. These modulations of form and color perception by motion signals must reduce the inconsistency of color and form with motion and thereby can be used to improve the visual quality produced by Deformation Lamps. In addition, the inconsistency between dynamic luminance edges and color edges may be resolved by form-color interactions \cite{AVV2012}. 

For large image distortions, Deformation Lamps also produces visible changes in static pattern including shape from shading. These static changes are often perceptually merged nicely with the original object. This is presumably because there are little image cues for the visual system to segment out the added components that produce only luminance changes in the final image.

Deformation Lamps fully utilizes the characteristics of the processing pipeline of the human visual system to analyze motion, color, and form separately and integrate them into coherent object representations.

\subsection{Experiment 1}
Our theory predicts two properties of Deformation Lamps that should be dependent on factors related to the human perceptual mechanism. For example, there must be a maximum limit to the physical magnitude of deformation that is allowed for the projected image, $I_{dynamic}$. Beyond a certain limit, the brain will be unable to resolve the inconsistency between the dynamic projection image and the static object. The purpose of this experiment was to evaluate the effects of changing the deformation magnitude, viewing distance, and deformation magnitude on the perceived image deformation of Deformation Lamps. 

\begin{figure}[!t]
  \centering
  \includegraphics[width=4.5in]{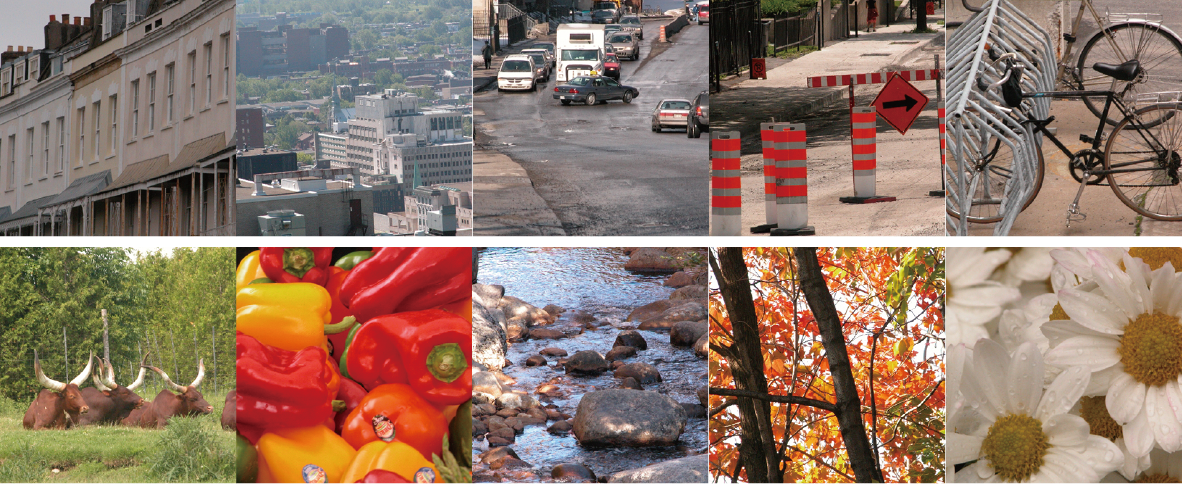}
  \caption{ Stimulus images that were used in Experiments 1 and 2. The five upper scenes contain man-made objects while the five lower contain only natural objects. All ten images were tested in Experiment 1, while the images on the left side of each column (i.e., one man-made scene and one natural scene) were tested in Experiment 2.}
\end{figure}

\subsubsection*{Method}
We used a video projector (EB-1761W, EPSON) to illuminate static pictures that were printed out with a laser printer (DocuCentreIV, C4475, Fuji Xerox). A computer (Mac Pro, Apple) controlled the stimulus presentation and data collection. Magnets were used to put the static pictures on a white board. 

We picked ten images of natural scenes (Figure 3) from the McGill Calibrated Colour Image Database \cite{OK2004}. Five scenes contained man-made objects while the rest contained only natural objects. The center square region of each image was cropped and printed out in the center of an A3-size sheet of paper. The size of the printed image was 13.2 $\times$ 13.2 cm. 

The deformation pattern was a horizontal sine wave.

\begin{equation}
D(x,y,t)=Asin(2πf_{s}y+\phi_{s} )cos(2πf_{t}t+\phi_{t}).
\end{equation}

We used six levels of spatial deformation amplitude ({\it A} = 0.1, 0.2, 0.4, 0.8, 1.7, and 3.3 cm). The spatial frequency of the deformation ($f_{s}$) was 1, 2, or 4 cycles per image. The spatial phase of the sinusoidal deformation ($\phi_{s}$) was randomized across images and observers. The temporal frequency of the deformation ({\it $f_{t}$}) was fixed at 1 Hz. The temporal phase of the deformation ($\phi_{t}$) was fixed at 0. The dynamic image sequence ($I_{dynamic}$) was produced from this distortion function and gray-scale images directly computed from the scene images, following (3). The projected image sequence was generated using (5). Here, $w$ was 0.4 and $B$ was a medium gray level. The value of $w$ was determined such that projected image sequences did not mask the appearance of the static picture while producing the greater effects on apparent picture deformation. Each stimulus movie consisted of a temporal alternation between one of the 1 sec luminance motion sequences and a 1 sec static uniform field that had the spatiotemporally averaged luminance of the luminance motion sequence (that is, $B$). When the uniform field was presented, only the static picture was illuminated, and thus it was seen as it is. When the luminance motion sequence was projected onto a static picture, whether or not the picture was perceived as deformed depended on the stimulus condition. 

Six people with normal or corrected-to-normal visual acuity participated as observers. They sat 110 cm or 220 cm from the white board on which a static picture was attached by magnets. In each trial, a dynamic image sequence was projected onto the static picture. The observers were asked to view the display and judge whether the static picture was seen as deforming or not. They reported their judgments by pressing the assigned keys of the keyboard of the computer on their lap. After their judgments were input, the next trial began. Each natural image and each observation distance was blocked in a single session consisting of 36 trials (3 deformation frequencies $\times$ 6 amplitude levels $\times$ 2 replications). For example, after performing the test with the 110 cm observation distance condition with one natural image picture, the observer  performed one with a 220 cm observation distance condition with the same natural image picture. After the observations with two observation distance conditions had been completed, an experimenter replaced the natural image picture with one of the untested pictures, and the next session started. Half of the observers performed the experiment in the opposite order of viewing distance conditions. We randomized the order of natural images to be evaluated across the observers. 

\begin{figure*}[!t]
  \centering
  \includegraphics[width=4.5in]{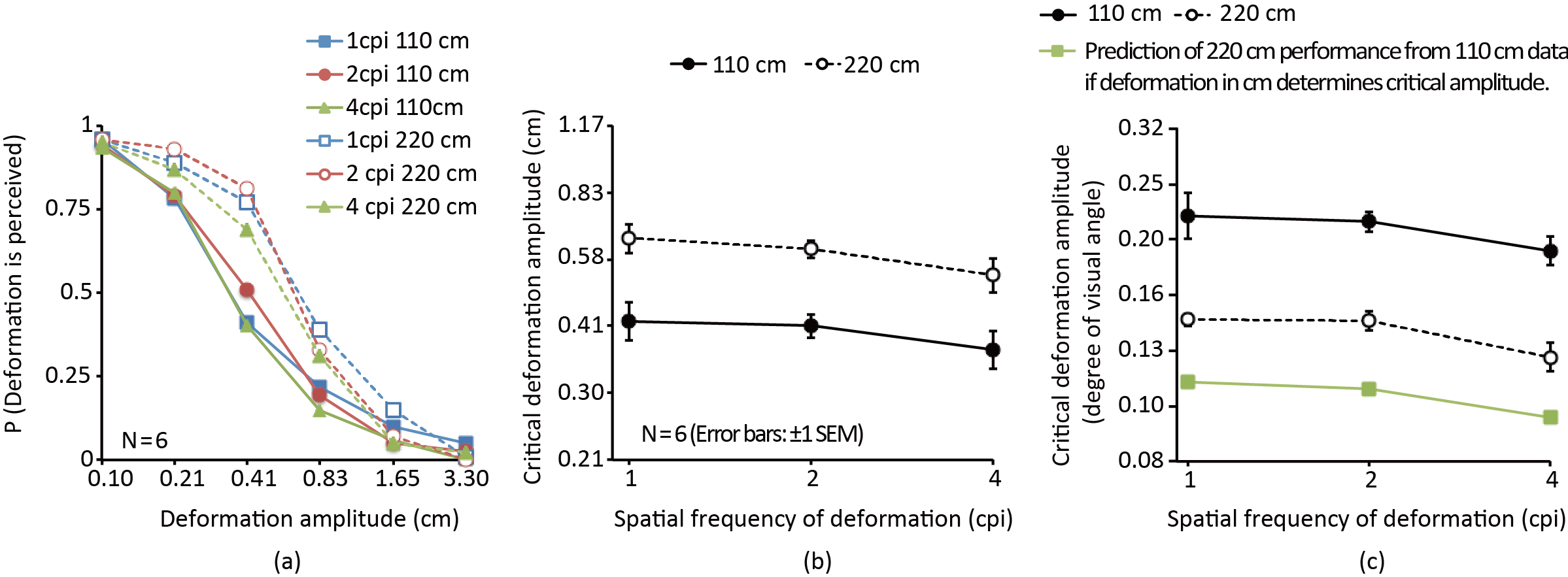}
  \caption{Experiment 1's results. (a) Proportions of the trials wherein the deformation of a static picture was seen plotted as a function of the deformation amplitude. (b) Critical deformation amplitude in cm. Error bars are within-subject standard errors of the mean \protect\cite{OC2014}. (c) Critical deformation amplitudes in degree of visual angle.}
\end{figure*}

\subsubsection*{Results and Discussion}

Plots with solid lines in Figure 4a show the proportions of trials in which the observers reported deformation as a function of the deformation amplitude. The viewing distance was 110 cm. As the deformation amplitude increased, the observers reported deformations less often. When they did not see the distortion, they saw the static picture and projected dynamic pattern to be segregated. We fitted a cumulative Gaussian function to the data of each observer and calculated critical deformation amplitudes below which more than 50$\%$ reports of the deformation were obtained. Plots with solid lines in Figure 4b show that regardless of the spatial frequency of deformation, the critical deformation amplitudes were about 0.4 cm. This implies that for an image viewed at a distance of 110 cm, Deformation Lamps can produce an apparent deformation effect when the deformation amplitude is below 0.4 cm. 

The critical deformation amplitude was dependent on the viewing distance, however. Plots with dashed lines in Figure 4a show the data in the trials in which the observers viewed a projection target from 220 cm away. In comparison with the 110 cm viewing distance, the function of the proportions in Figure 4a shifted rightward, and the critical deformation amplitude was significantly greater at 220 cm than at 110 cm (Figure 4b). We conducted a two-way repeated measures analysis of variance with viewing distance and spatial frequency of deformation as factors, and found that the main effect of viewing distance was significant [{\it F}(1,5) $=$ 22.013, {\it p} $<$ .006]. Thus, the critical deformation amplitude was not determined by the physical amplitude of the deformation on the display. On the other hand, the critical deformation magnitude was not determined by the deformation amplitude in terms of visual angle (i.e., the deformation magnitude on the retina), either. As shown in Figure 4b, the critical deformation amplitude in visual angle was significantly smaller at 220 cm than at 110 cm. This is likely due to the change in the overall retinal image size in accordance with the viewing distance. In sum, Deformation Lamps can generate perceptual deformations as long as the deformation amplitude is small, and the critical deformation amplitude beyond which the perceptual distortion collapses is 0.4-0.6 cm on the display or 0.2-0.3 deg in visual angle, at least in the situations we tested.

\subsection*{Experiment 2}
The first experiment demonstrated that Deformation Lamps can generate perceptual deformations for a range of small distortion amplitudes. When this happens, does the observer perceive the distortion magnitude as large as that of the distortion of the projected pattern? We suggested that how much perceptual deformation Deformation Lamps actually produces may not be predictable solely from the physical magnitude given to $I_{dynamic}$. Considering the inconsistency of the motion signals with the static pattern and color information, it is likely that the perceived distortion magnitude is somehow reduced, but if so, how much is the reduction? This is an important point to know for the sake of image quality control. The second experiment thus evaluated the magnitude of distortion perceived in images treated with Deformation Lamps in comparison with the perceived magnitude of real physical distortions.

\begin{figure}[!t]
  \centering
  \includegraphics[width=4.5in]{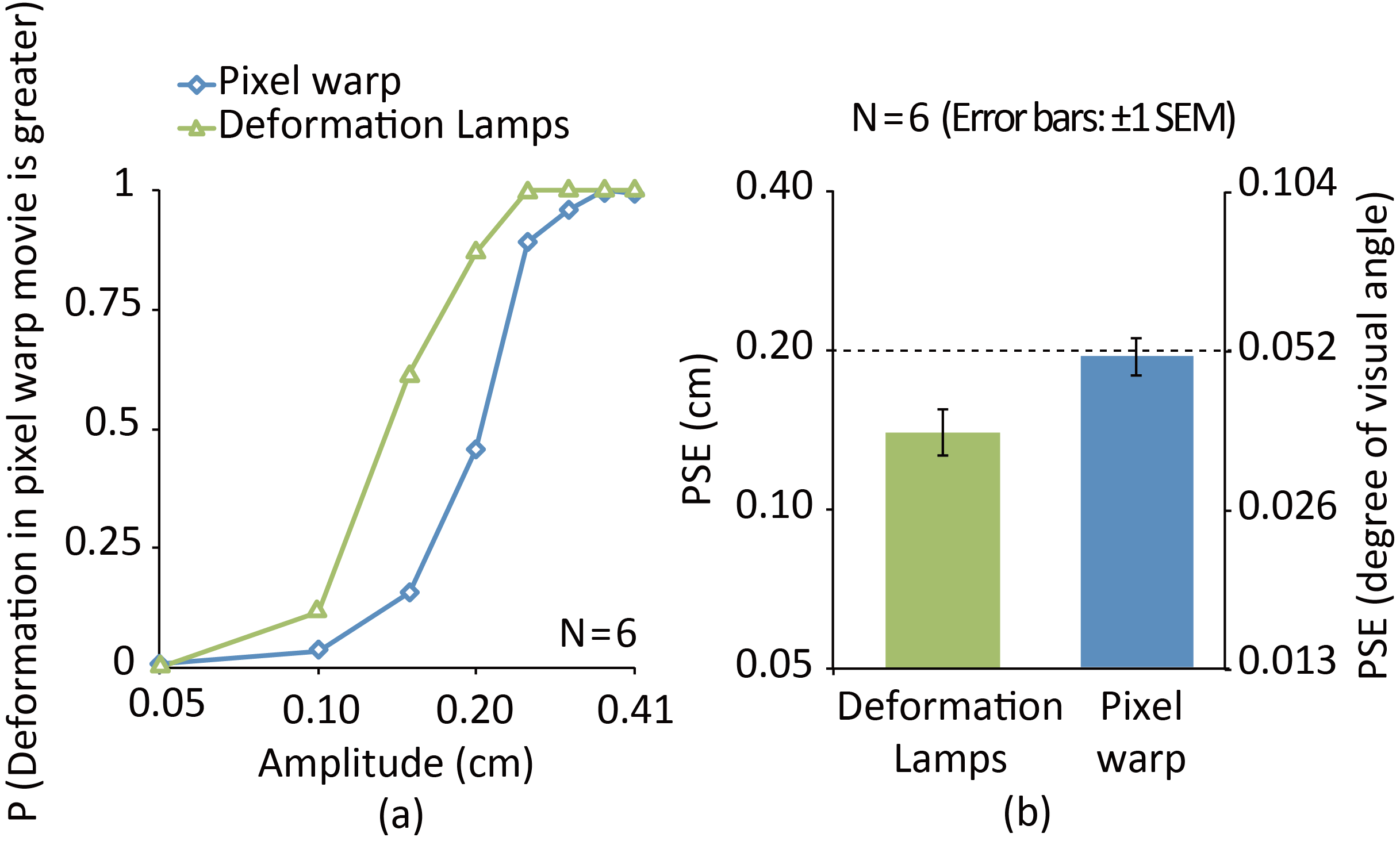}
  \caption{Experiment 2's results. (a) The proportions of trials in which the deformation in the right-side movie was reported to be greater than that in the left-side movie. (b) The point of subjective equality (PSE) of deformation between the left and right movies for the Deformation Lamps and Pixel warp conditions. }
\end{figure}

\subsubsection*{Method}
The observers and apparatus were those in Experiment 1. In each trial, two deformation movies were simultaneously presented on the left and right sides of a paper screen. The right side of the movies was either a pixel warp (physical distortion) movie on a white blank screen (Pixel warp condition) or a dynamic luminance movie on a static picture (Deformation Lamps condition). The deformation amplitude was 0.21 cm, and the viewing distance was 110 cm. The left side of the movies was a pixel warp movie in which the amplitude of the deformation was one of eight alternative magnitudes (0.05, 0.1, 0.15, 0.21, 0.26, 0.31, 0.36, and 0.40 cm). The task of the observers was to judge which of the two movies apparently had a larger deformation. 

\subsubsection*{Results and Discussion}

We calculated the proportion of trials in which the movie on the left side was reported to have the greater amplitude of deformation and plotted the proportion as a function of the deformation amplitude (Figure 5a). We fitted a cumulative Gaussian function to the data and calculated a point of subjective equality (PSE) for the perceived deformation magnitudes (Figure 5b). This was the point of the spatial amplitude at which the right and left movies produced subjectively equal magnitudes of deformation. The results demonstrated that the PSE was significantly smaller under the Deformation Lamps condition than under the Pixel warp condition [\textit{t}(5) = 3.40, \textit{p} $<$ .02]. The mean PSE in the Pixel warp condition was 0.199 cm, and this indicates that the observers of this experiment could make veridical comparisons of the deformation amplitudes, because the actual amplitude of the deformation in the right-side movie was 0.21 cm. On the other hand, the mean PSE in the Deformation movie conditions was 0.144 cm, and this indicates that the observers' perception of the deformation amplitudes was underestimated.

The results suggest that the magnitude of deformation produced by Deformation Lamps on a static picture is not as much as the physical deformation of the projected image, by $\sim$30$\%$ under our tested conditions. It is likely that the rate of reduction is not a constant value, but changes depending on a number of conditions, including the distortion magnitude and projection image contrast.

\section{Applications}

\subsection{Picture deformation by camera-projection system}
The camera-projection system, described in Figure 1, can produce apparent dynamic deformation in a variety of static objects, including drawn pictures, printed photographs, sculptures with 3D shading, objects with natural textures (e.g., vegetables) and even human bodies.

Image deformations give observers a variety of impressions, since many physical factors in the real world produce image distortions, and the deformation pattern contains information about the physical cause. Sometimes, the object itself physically deforms in a specific way. Sometimes the image deformation of the object is produced by light refracting through transparent material lying between the object and the observer, such as air and water, and the pattern of deformation reflects the fluid dynamics of the material. The image deformation due to refraction is a strong cue of the presence of transparent materials for computer vision \cite{X2014}. For human vision, the impression of transparent liquid is generated solely by dynamic image deformation, and the deformation pattern could be a random modulation of a specific range of spatiotemporal frequencies \cite{KMN2015}.

Taking advantage of knowledge about natural image distortions, Deformation Lamps can make a picture of flames fluttering in the dark, a picture of a stone under moving water (Figure 6), and a picture of a road scene with a waving heat shimmer. In these cases, the deformation map can be defined independently of the image content. Thus, the computation of the projected dynamic image sequence is light and quick. For more precise control of the deformation pattern, Deformation Lamps could be combined with such image editing techniques, as proposed by \cite{C2005}, and \cite{BBBVN2011}. 

\begin{figure*}[!b]
  \centering
  \includegraphics[width=4.5in]{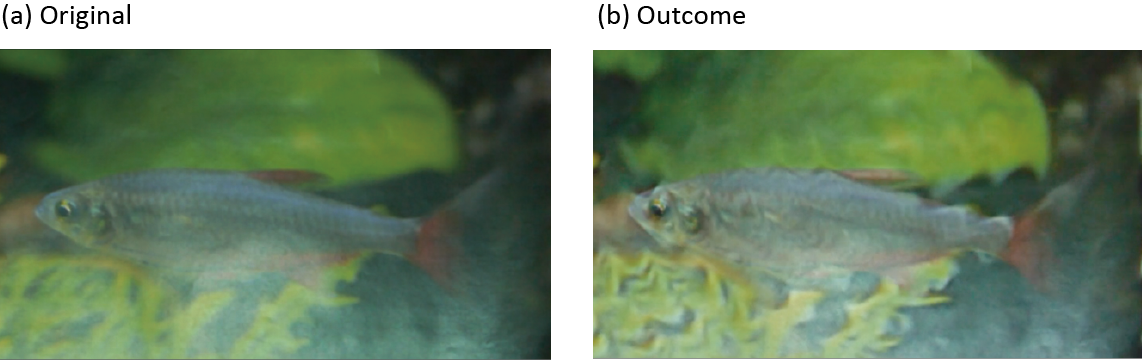}
  \caption{An example of under-water effect. (a) Original photograph of fish. (b) Outcome of our technique. In (b), we perceive as if fish is under water. The under-water effect is more pronounced in a movie than in a still image. See Movie 1 to check it.}
\end{figure*}

\begin{figure*}[t]
  \centering
  \includegraphics[width=4.5in]{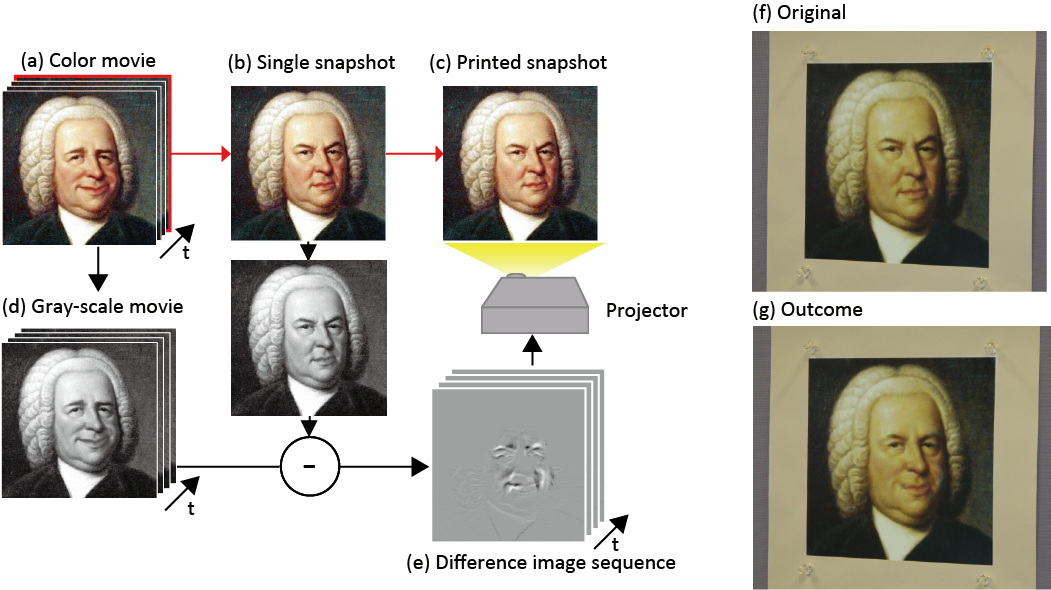}
  \caption{Processing stream of picture-based movie.  (a) A color movie the system attempts to reproduce. (b) A single snapshot extracted from (a). (c) A printout of (b) set as a projection target. (e) A difference image sequence made from subtraction of (e) from (d). This image sequence (e) is projected onto (f) to produce a movie that looks like (g) to human observers. }
\end{figure*}

\subsection{Picture-based movie}

Instead of using a general-purpose deformation maps, one can design an image distortion specific to the target object, e.g., facial expression to a portrait. This can be done by making a movie from a static picture by hand (using Photoshop) or by using special software. Alternatively, while the camera-projection system produces a movie from a target picture, we can reverse this relationship and produce a picture from an existing movie. 

Let us return to (2) and (3). Now $I_{movie}$ is the movie we want to show, and $I_{static}$ is a static two-dimensional image. $I_{static}$ is printed out in color, and $I_{dynamic}$ computed by (3) is projected onto the picture with an arbitrary gray background. Due to the limitations revealed by the psychophysical experiments, the amount of deformation of each frame relative to the key frame should not be large. For this reason, it is better to choose a snapshot image close to the sequence-averaged image. As we expected, by projecting the luminance dynamic image onto a printed snapshot image of the movie (Figure 6), the observer had the impression of an original-movie-like motion and deformation. This technique is able to add natural biological motion to a static dead object, enabling one to make something akin to Harry Potter moving pictures.

\subsection{Object deformation by transmissive LCD}
The basic idea of Deformation Lamps is to deform static objects by addition of dynamic luminance component. We can apply this idea to other means of image mixture, such as a transmissive liquid crystal display (LCD) (Figure 5). We confirmed that by looking at a 3D object through the transmissive LCD on which a dynamic image sequence is presented (Figure 7), the observers could indeed perceive deformation of the object. This method can apparently deform even the boundaries of the object. It is not affected by the surface reflectance property of the object at all. In addition, it can be used even where there is no space for a projection setup.

One disadvantage of a transmissive LCD however is that the viewpoint is severely limited as the object has to be aligned behind the transmissive LCD with the dynamic image sequence on the display. Another problem is a brightness reduction because light significantly diminishes in intensity when it is transmitted through the LCD. 

\begin{figure}[!ht]
  \centering
  \includegraphics[width=4.5in]{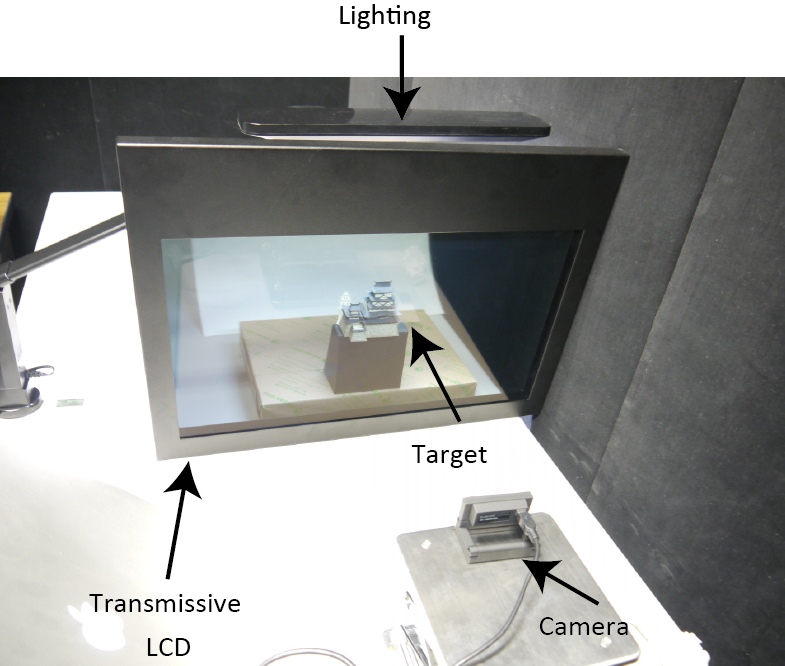}
  \caption{Addition of perceptual deformations to a 3D object by a transmissive LCD.}
\end{figure}

\section{Limitations and future issues}
Deformation Lamps cannot add the impression of enduring unidirectional motion to a static object. As shown in experiment 1, the spatial coordinates of the luminance motion signals need to match those of the static object. If enduring unidirectional motion is presented, the consistency of the spatial coordinates between luminance motion signals and the static object will be quickly violated, and the observer will lose the dynamic impression of a static object. For the same reason described above, Deformation Lamps is not suitable when one wants to give a dynamic impression with large speed motion vectors to a static object.

The present study consistently used a static object as a projection target. But in principle, it is possible to give dynamic impressions to the surface of a moving object. Using Deformation Lamps on moving objects will be a challenging task. For example, one difficulty to run the camera-projection system (Figure 1) to a moving object is how to get a sequence of gray-scale image of the undistorted projection target.  

Combining Deformation Lamps with other light projection techniques may provide an effective way to change both static and dynamic aspects of the appearance of a real object. For instance, Deformation Lamps with a camera-projection system can add motion impressions to a static target object whose color, texture and surface material are simultaneously modified by Shader Lamps \cite{RWL2001}. In this system, the Shader Lamps computation does not have to consider the object's movements at all if the Deformation Lamps computation can access the object appearance modified only by Shader Lamps. As such, the two techniques can be combined very easily.

The image quality of Deformation Lamps is significantly affected by a variety of factors such as reflection characteristics of the object surface, stimulus contrasts of the projecting and projected images, and the position of the viewer. For practical application of Deformation Lamps, we need to know more about such things as printing materials, printing methods and projection systems suited for this technique. Even under non-ideal conditions, the system can calibrate the projected image to cancel the effects of various image distortion factors, and obtain the desired dynamic image.

Finally, production of illusory motion of a static object by addition of a luminance dynamic pattern described in (2) and (3) is a very general idea. As demonstrated by a transmissive LCD system (Figure 7), it can be realized in a variety of methods of image mixture other than light projection. Also, this idea may be applicable not only to dynamic image visualization, but also to effective compression of color movies.

\bibliographystyle{plain}
\bibliography{Arxiv_Kawabe}

\begin{thebibliography}{10}

\bibitem{AB1985}
E.~H. Adelson and J.~R. Bergen.
\newblock Spatiotemporal energy models for the perception of motion.
\newblock {\em Journal of the optical society of America A}, 2:284--299, 1985.

\bibitem{ALY2008}
D.~G. Aliaga, A.~Law, and Y.~H. Yeng.
\newblock A virtual restoration stage for real-world objects.
\newblock {\em ACM Transactions on Graphics}, 27(5):149:1—--149:10, 2008.

\bibitem{AYL2012}
D.~G. Aliaga, Y.~H. Yeng, and A.~Law.
\newblock Fast high-resolution appearance editing using superimposed
  projections.
\newblock {\em ACM Transactions on Graphics}, 31(2):13:1—--13:13, 2012.

\bibitem{AMANO2013}
T.~Amano.
\newblock Projection based real-time material appearance manipulation.
\newblock In {\em Proceedings of the 2013 IEEE Conference on Computer Vision
  and Pattern Recognition Workshops}, pages 918--923, 2013.

\bibitem{AVV2012}
S.~Anstis, M.~Vergeer, and R.~Van~Lier.
\newblock Luminance contours can gate afterimage colors and ``real'' colors.
\newblock {\em Journal of Vision}, 12(10):2:1--13, 2012.

\bibitem{ANSTIS1970}
S.~M. Anstis.
\newblock Phi movement as a subtraction process.
\newblock {\em Vision Research}, 10:1411--1430, 1970.

\bibitem{AR1975}
S.~M. Anstis and B.~J. Rogers.
\newblock Illusory reversals of visual depth and movement during changes in
  contrast.
\newblock {\em Vision Research}, 15:957--961, 1975.

\bibitem{AOSS2006}
M.~Ashdown, T.~Okabe, I.~Sato, and Y.~Sato.
\newblock Robust content-dependent photometric projector compensation.
\newblock In {\em Proceedings of IEEE International Workshop on
  Projector-Camera Systems}, pages 60--67, 2006.

\bibitem{BBBVN2011}
M.~A. Batista, G.~Buscaglia, C.~Z. Barcelos, L.~Velho, and L.~G. Nonato.
\newblock Animating liquids in a still image.
\newblock In {\em Proceedings of Computer Graphics International 2011}, 2011.

\bibitem{BBGIBG2013}
A.~Bermano, P.~Br{\"u}schweiler, A.~Grundh{\"o}fer, D.~Iwai, B.~Bickel, and
  M.~Gross.
\newblock Augmenting physical avatars using projector-based illumination.
\newblock {\em ACM Transactions on Graphics}, 32:189:1--—189:10, 2013.

\bibitem{BEK2005}
O.~Bimber, A.~Emmerling, and T.~Klemmer.
\newblock Embedded entertainment with smart projectors.
\newblock {\em Computer}, 38(1):48--55, 2005.

\bibitem{BI2008}
O.~Bimber and D.~Iwai.
\newblock Superimposing dynamic range.
\newblock {\em ACM Transactions on Graphics}, 27(5):150:1--150:8, 2008.

\bibitem{BR2005}
O.~Bimber and R.~Raskar.
\newblock {\em Spatial augmented reality. Merging real and virtual worlds.}
\newblock A K Peters, Wellesley, Massachusetts, 2005.

\bibitem{BIWG2008}
Oliver Bimber, Daisuke Iwai, Gordon Wetzstein, and Anselm Grundh{\"o}fer.
\newblock The visual computing of projector-camera systems.
\newblock In {\em ACM SIGGRAPH 2008 Classes}, SIGGRAPH '08, pages 84:1--84:25,
  2008.

\bibitem{BT2011}
D.~Burr and P.~Thompson.
\newblock Motion psychophysics: 1985-2010.
\newblock {\em Vision Research}, 51:1431--1456, 2011.

\bibitem{CTF1984}
P.~Cavanagh, C.~W. Tyler, and O.~E. Favreau.
\newblock Perceived velocity of moving chromatic gratings.
\newblock {\em Journal of the optical society of America A}, 1:893--899, 1984.

\bibitem{CLQW2008}
M.~Chi, T.~Lee, Y.~Qu, and T.~WONG.
\newblock Self-animating images: Illusory motion using repeated asymmetric
  patterns.
\newblock {\em ACM Transactions on Graphics}, 27(3):62:1--62:8, 2008.

\bibitem{C2005}
Y-Y. Chuang, D.~B. Goldman, K.~C. Zheng, B.~Curless, D.~H. Salesin, and
  R.~Szelski.
\newblock Animating pictures with stochastic motion textures.
\newblock {\em ACM Transactions on Graphics}, 24(3):853--860, 2005.

\bibitem{CW2005}
S.~J. Cropper and S.~M. Wuerger.
\newblock The perception of motion in chromatic stimuli.
\newblock {\em Behavioral and Cognitive Neuroscience Reviews}, 4:192--–217,
  2005.

\bibitem{DD1991}
R.~L. De~Valois and K.~K. De~Valois.
\newblock Vernier acuity with stationary moving gabors.
\newblock {\em Vision Research}, 31:1619–--1626, 1991.

\bibitem{D1929}
K.~Duncker.
\newblock Uber induzierte bewegung.
\newblock {\em Psychologische Forschung}, 12:180--259, 1929.

\bibitem{FW1979}
A.~Fraser and K.~J. Wilcox.
\newblock Perception of illusory movement.
\newblock {\em Nature}, 281, 1979.

\bibitem{FAH1991}
W.~Freeman, E.~H. Adelson, and D.~J. Heeger.
\newblock Motion without movement.
\newblock In {\em Proceedings of ACM SIGGRAPH `91}, pages 27--30, 1991.

\bibitem{KMN2015}
T.~Kawabe, K.~Maruya, and S.~Y. Nishida.
\newblock Perceptual transparency from image deformation.
\newblock {\em Proceedings of the National Academy of Sciences}, Early edition,
  2015.

\bibitem{K1980}
D.~Kelly.
\newblock Motion and vision. ii. stabilized spatio-temporal threshold surface.
\newblock {\em Journal of the Optical Society of America}, 69:1340--1349, 1980.

\bibitem{KA2003}
A.~Kitaoka and H.~Ashida.
\newblock Phenomenal characteristics of the peripheral drift illusion.
\newblock {\em Vision (Journal of the Vision Society of Japan)}, 15:261--262,
  2003.

\bibitem{LBJ2004}
R.~Legarda-S{\'a}enz, T.~Bothe, and W.~P. J{\"u}ptner.
\newblock Accurate procedure for the calibration of a structured light system.
\newblock {\em Opt. Eng.}, 43(2):464--471, 2004.

\bibitem{LH1985}
M.~S. Livingstone and D.~H. Hubel.
\newblock Spatial relationship and extrafoveal vision.
\newblock {\em Nature}, 315:285, 1985.

\bibitem{LH1987}
M.~S. Livingstone and D.~H. Hubel.
\newblock Psychophysical evidence for separate channels for the perception of
  form, color, movement, and depth.
\newblock {\em Journal of Neuroscience}, 7:3416--3468, 1987.

\bibitem{LS1995}
Z.~L. Lu and G.~Sperling.
\newblock The functional architecture of human visual motion perception.
\newblock {\em Vision Research}, 35:2697–--2722, 1995.

\bibitem{MNS2004}
Y.~Mukaigawa, M.~Nishiyama, and T.~Shakunaga.
\newblock Virtual photometric environment using projector.
\newblock In {\em Proceedings of the International Conference on Virtual
  Systems and Multi- media}, pages 544--553, 2004.

\bibitem{NISHIDA2004}
S.~Nishida.
\newblock Motion-based analysis of spatial patterns by the human visual system.
\newblock {\em Current Biology}, 14:830--839, 2004.

\bibitem{NISHIDA2011}
S.~Nishida.
\newblock Advancement of motion psychophysics: Review 2001-2010.
\newblock {\em Journal of Vision}, 11(5):1--53, 2011.

\bibitem{NJ1999}
S.~Nishida and A.~Johnston.
\newblock Influence of motion signals on the perceived position of spatial
  pattern.
\newblock {\em Nature}, 397:610--612, 1999.

\bibitem{NWKT2007}
S.~Nishida, J.~Watanabe, I.~Kuriki, and T.~Tokimoto.
\newblock Human visual system integrates color signals along a motion
  trajectory.
\newblock {\em Current Biology}, 17:366--372, 2007.

\bibitem{OC2014}
F.~O'Brien and D~Cousineau.
\newblock Representing error bars in within-subject designs in typical software
  packages.
\newblock {\em The Quantitative Methods for Psychology}, 10:56--67, 2014.

\bibitem{OK2004}
A.~Olmos and F.~A.~A. Kingdom.
\newblock A biologically inspired algorithm for the recovery of shading and
  reflectance images.
\newblock {\em Perception}, 33:1463--1473, 2004.

\bibitem{RAMA1987}
V.~S. Ramachandran.
\newblock Interaction between colour and motion in human motion.
\newblock {\em Nature}, 328:645--647, 1987.

\bibitem{RAMAAN1990}
V.~S. Ramachandran and S.~M. Anstis.
\newblock Illusory displacement of equiluminous kinetic edges.
\newblock {\em Perception}, 19:611--616, 1990.

\bibitem{RAMACAVA1987}
V.~S. Ramachandran and P.~Cavanagah.
\newblock Motion capture anisotropy.
\newblock {\em Vision Research}, 27(1):97--106, 1987.

\bibitem{RAMAGRE1978}
V.~S. Ramachandran and R.~L. Gregory.
\newblock Does colour provide an input to human motion perception?.
\newblock {\em Nature}, 275:55--56, 1978.

\bibitem{RWF1998}
R.~Raskar, G.~Welch, and H~Fuchs.
\newblock Spatial augmented reality.
\newblock In {\em Proceedings of the First IEEE Workshop on Augmented Reality
  (IWAR'98)}, pages 1--7, 1998.

\bibitem{RWL2001}
R~Raskar, G~Welch, and K-L Low.
\newblock Shader lamps: Animating real objects with image-based illumination.
\newblock In {\em Proceedings of the 12th Eurographics Workshop on Rendering
  Techniques}, pages 89--102, 2001.

\bibitem{RZW2002}
R.~Raskar, R.~Ziegler, and T.~Willwacher.
\newblock Cartoon dioramas in motion.
\newblock In {\em Proceedings of the 2nd international symposium on
  Non-photorealistic animation and rendering}, pages 7--12, 2002.

\bibitem{U1997}
J.~Underkoffler.
\newblock A view from the luminous room.
\newblock {\em Personal Technologies}, 1:49--59, 1997.

\bibitem{VS1985}
J.~P. Van~Santen and G.~Sperling.
\newblock Elaborated reichardt detectors.
\newblock {\em Journal of the optical society of America A}, 2:300--320, 1985.

\bibitem{WANDELL1995}
B.~A. Wandell.
\newblock {\em Foundations of Visual Science}.
\newblock Sunderland, Massachusetts: Sinauer Associates, 1995.

\bibitem{X2014}
T.~Xue, M.~Rubinstein, N.~Wadhwa, F.~Durand, and W.~T. Freeman.
\newblock Refraction wiggles for measuring fluid depth and velocity from video.
\newblock In {\em Proceedings of European Conference on Computer Vision
  (ECCV)}, pages 767--782, 2014.

\end{thebibliography}

\end{document}